
\magnification=1200
\hsize 16.0truecm
\vsize 22.0truecm
\voffset=-1. truecm         
\baselineskip=12pt
\parskip=5pt
\parindent=22pt
\raggedbottom



\def\pp{\noindent\parshape 2 0.0 truecm 17.0 truecm 0.2 truecm 16.5 truecm}

\font\bigbf= cmb10 scaled\magstep2

\def\pn{\par\noindent}

\def\etal{{\it et~al.~}}
\def\dv{$\Delta V_{HB}^{TO}$~}

\def\bv0g{$(B-V)_{0,g}$~}
\def\lz{$(B-R)/(B+V+R)~$}

\def\fe{[Fe/H]~}

\pn
\null\vskip 5.5truecm

\centerline{\bf THE YOUNG GLOBULAR CLUSTERS OF THE MILKY WAY}
\par\noindent
\centerline{\bf AND THE LOCAL GROUP GALAXIES:}
\par\noindent
\centerline{\bf  PLAYING WITH GREAT CIRCLES}
\par\noindent
\bigskip
\bigskip
\centerline {\bf F. Fusi Pecci$^1$, M. Bellazzini$^2$, C. Cacciari$^1$,
F.R. Ferraro$^1$}
\bigskip
\bigskip
\centerline {\it $^1$ Osservatorio Astronomico$^*$, Bologna, Italy}
\pn
\centerline {\it $^2$ Dipartimento di Astronomia$^*$, Bologna, Italy}
\pn
\centerline {\it $^*$V. Zamboni 33, I-40126, Bologna, Italy}
\bigskip
\bigskip
\centerline{BAP 07-1995-032-OAB}
\bigskip
\bigskip
\bigskip
\vskip 2 true cm
\centerline{\it The Astronomical Journal, in press.}
\vskip 2 true cm
{\it Send proofs to:} M. Bellazzini
\par
E-mail: bellazzini@astbo3.bo.astro.it

\vfill\eject
\noindent
\bigskip\bigskip

\centerline {\bf ABSTRACT}
\bigskip\noindent
The small group of Galactic Globular Clusters (GGC) (Pal 12, Terzan 7,
Ruprecht 106, Arp 2) recently discovered to be significantly younger
(by $\sim 3-4$ Gyr) than the average cluster population of the Galaxy are
shown to lie near planes passing in the vicinity of some satellite
galaxies of the Milky Way and through the Galactic Centre itself.

Assuming that these configurations represent a fossil record of interactions
between the Galaxy and its companions from which these clusters originated,
we identified, along one of them, another candidate ``young'' GGC,
{\it i.e.} IC4499, whose Color-Magnitude Diagram is presented.

Various hypotheses on the possible  origin of ``young'' GGC are also
briefly discussed within a framework where the
location on preferential planes may be seen as a general characteristic
for the Local Group members.

\vfill\eject

\noindent
{\bf 1. INTRODUCTION}

\medskip\noindent
The evidence that several satellites of the Milky Way appear to be situated
along a few great circles (streams), and that this may be somehow
related to their origin and dynamical and chemical evolution, has been
a sort of unsettled but persistent theme in the last twenty years or
so for most studies of the stellar systems (galaxies and clusters)
populating the Local Group.
In particular, Hodge and Michie (1969), Kunkel and Demers (1977),
Kunkel (1979), Lynden-Bell (1976, 1982) and others (see the reviews by
Majewski 1993a,b)
pointed out the existence of two remarkable great circles which account
for most of the dwarf satellites of the Milky Way.
The first, {\it the Magellanic Plane} (MP, Kunkel and Demers 1977), at an
angle of about 40$^o$ with respect to the plane defined by the Magellanic
Stream (MS), roughly passes  through the Magellanic Clouds,
the Galactic Centre and the Draco-Ursa Minor region. The second, called by
Lynden-Bell (1982) {\it the Fornax - Leo - Sculptor Stream} (FLS), ideally
contains in a plane Fornax, Leo I and II, and Sculptor.
\par
Concerning the Galactic globular clusters (GGCs), as recently reviewed and
discussed for instance by Zinn (1993), van den Bergh
(1993), and Majewski (1993b, 1994), there are now many indications suggesting
the existence of different populations (e.g. disk, old and young halo, etc.)
possibly originated via different mechanisms.
Moreover, the possibility of an early history of the Galaxy affected by
tidal interactions, mergers or captures, involving satellite galaxies
and outer halo globular clusters has recently received a noticeable support
by the latest discovery of the Sagittarius dwarf spheroidal
galaxy (Sgr dSph) currently being disrupted and absorbed by the Milky Way
(Ibata \etal 1994, Mateo \etal 1994).
\medskip
Our specific interest in this subject originates from the early detections
of a small (but growing) set of globular clusters which are
significantly younger (by 3-4 Gyr) than the bulk of the cluster
population studied so far in our own Galaxy, {\it i.e.} Pal 12 (Gratton and
Ortolani 1988, Stetson \etal 1989), Ruprecht 106 (Buonanno \etal 1990, 1993),
Arp 2 (Buonanno \etal 1994a,b), and Terzan 7 (Buonanno \etal 1994a,c).
As stressed in a previous paper (Buonanno \etal 1994a), we
noted that these four clusters appear to lie nearly along a great circle in
the sky and, following an early suggestion by Lin and Richer (1992),
we were led to conclude that they could be on similar orbits and
may have been captured by the Milky Way. Our present analysis of the
available data on locations and kinematics of the known stellar
systems in the Local Group is strongly centered on the problem
of the origin of the young globulars of the Galaxy and has two aims:
\pn
(1) Search for additional candidate young Galactic globular clusters among
those which are suitably located with respect to the most attractive
configurations identified between the four known young globulars and
the satellite galaxies.
\pn
(2) Analyze the configurations somehow linking young GGCs
to the other satellites of the Milky Way to see if and how they
could fit into the picture of a few great circles (planes)
representing the signature of recent or old connections.
\par
This is only a first step towards a comprehensive study of the dynamics,
kinematics and chemical enrichment history of these stellar components, which
will require more data (in particular proper motions and detailed chemical
abundances) before it can produce conclusive results.
However we believe it is worth reporting some preliminary results and
speculations as a possible guide for further observations and models,
especially since these have led us to pick up, maybe fortuitously,
another ``young candidate'' ({\it i.e.}
IC 4499, Ferraro \etal 1994), which confirms the potentiality of such
a procedure.
\bigskip\noindent
\noindent
{\bf 2. DATA-SET AND PROCEDURE}

\medskip\noindent
Our database consists of all the Local Group galaxies, and the GGCs
more distant than 10 Kpc from the Galactic Centre.
The main reason for this choice is that since the Galactic
Centre is the pivot of any plane, the objects too close to that point
do not provide any significant information to distinguish between
different solutions.
The basic data-sources for the GGCs are from
Thomas (1989, coordinates and radial velocities), Djorgovski
(1993, metallicities), Armandroff (1989) and Peterson (1993)
for $V_{HB}$ and $E(B-V)$. If missing from these sources, radial velocities
and metallicities
have been taken from other lists (Zinn 1985, Webbink 1985,
Armandroff and Da Costa 1991, Freeman \etal 1983).
The heliocentric and galactocentric distances have been calculated with the
same assumptions as Armandroff (1989).
The data on Local Group galaxies are drawn from van den Bergh
(1994a) and Zaritsky (1994).
It is important to stress that the use of any other reference source for
these parameters would make substantially no difference since
we are just looking at overall configurations hardly affected by
the uncertainties associated to these observables.
Conversely, it is important to note that the adopted distance scale
is strictly homogeneous, so the description of the GGC system is fully
self-consistent as far as standard candles and zero-points
are concerned.
\par
To carry out our tests we have selected various groups of interesting objects
(for instance LMC, SMC, all dSph's, Arp 2, Terzan 7, Ruprecht 106, Pal 12 and
many other objects in
some way connected to them in previous studies) and have interactively
checked their locations
with respect to planes passing through any pair of them and the Galactic
Center. When an interesting configuration was found we simply checked
the characteristics of the possible members and then optimized
the matching of the parameters involved in the description.
The criteria that guide our judgement in recognising an ``interesting''
configuration are that a) the configuration is clearly defined, i.e. the
distances of each object from the defined plane are as small as possible;
and b) the maximum number of objects, in particular of young GGCs and Galaxy
satellites, belong to the configuration. The position of the projection in the
sky of the considered plane with respect to the Magellanic Stream was also
considered. See Sect. 3.1 for more details.
\pn
The diagrams presented in Figure 1 and 2 are made adopting the
approach of Lynden-Bell (1982,
Table I), {\it i.e.} we plot {\it (a)} the coordinates $l_{G}$, $b_{G}$
at which each object would be seen if viewed from the Galactic Centre,
assumed to be at 8 Kpc from the sun and {\it (b)} the Galactic Great Circles
which are the projection on the sky of the main planes discussed in text.
The approximate location and extension of the Magellanic Stream has also
been reported as derived from Kunkel (1979, Fig. 1).

\bigskip\noindent
{\bf 3. RESULTS}

\bigskip\noindent
Figure 1 shows a diagram where {\it all} the objects (clusters and
galaxies) considered in the present analysis are plotted in the
adopted reference frame ($l_{G}$, $b_{G}$) together with the Galactic Great
Circles representing the quoted Magellanic Plane and
Fornax-Leo-Sculptor Plane. If one considers that these
planes are also passing through the Galactic Centre, some alignments are
at least curious. For instance, the newly discovered Sgr dSph is not far from
the Magellanic Plane in the sky, and the FLS-plane accounts fairly consistently
for Sextans and even Phoenix (at about 400 Kpc indeed, see Majewski 1994).
However, one has to recall that the Magellanic Clouds and the pair Draco-Ursa
Minor lie nearly at the antipodes on the sky, and
keeping these two {\it nodes} fixed it is possible to obtain
{\it a family} of Magellanic Planes. This is why
Kunkel and Demers (1977) and Lynden-Bell (1976) independently
found similar planes accounting for the MCs and the pair Draco - Ursa Minor,
yet tilted by an angle of more than 30$^o$.
On the other hand, though
a causal connection implying a tight alignment of MCs, Galactic Centre,
and the Draco-Ursa Minor pair is difficult to prove, it seems unlikely that
several other satellites would fit these configurations just by chance.
\par
Moreover, as pointed out by various authors (see Hernquist and Bolte 1993, and
references therein) if some globular clusters formed or were captured during
merger events  or galaxy interactions,  it seems natural that they would be
found in tails,
ribbon-like bridges and similar sub-structures which could retain their
dynamical identity for more than one Gyr (Johnston and Hernquist 1993,
Piatek and Pryor 1993, Johnston, Spergel and Hernquist 1995),
that is for a time of the order of the estimated orbital period of the
Sgr dSph (Velasquez and White 1995). Hence, one could in principle even
detect various different planar distributions of remnants
(though possibly deformed by later evolution) as a result of repeated
close encounters or of the orbital decay of a satellite
merging with a disk galaxy on a rather inclined orbit (see Quinn \etal 1993,
Fig. 11).

\medskip\noindent
{\bf 3.1. How many Magellanic Planes?}

\medskip\noindent
Following the procedure described in Sect. 2, we found several Magellanic
Planes which are compatible with many interesting objects.
As stressed by Lynden-Bell (1982), any two points in the sky have a
great circle passing through them, and one cannot pay too much
attention to all curious alignments. However, we note that two of
these great circles, which are shown in Figure 2, are particularly
interesting as they could also contribute to
shed some light on the possible origin of the young globulars.

\pn
The first plane, that we have named MP-1, is reported in Figure 2 as
a {\it full line}. In our view, the main characteristics that
make this configuration worth of consideration are:
\pn
(1) It contains the MCs, Draco-Ursa Minor and the two young globulars Pal 12
and Ruprecht 106, which were found by Lin and Richer (1992) to have kinematical
properties compatible with a capture from the MCs. Furthermore, van den Bergh
(1994b) has argued for the association of Pal 12 specifically with SMC, on
the basis of chemical and structural similarities.
\pn
(2) Its projection on the sky, {\it i.e.} the corresponding Great Circle,
is nearly superposed to the Magellanic Stream.
\smallskip\noindent
To define a general criterium of membership of any object to a given plane
we consider the angle formed by its radius vector to the Galactic Centre
and the plane itself, and assume membership if the probability for
this angle to be smaller than presently observed in a random distribution
is less than $15 \%$.
The candidate members of MP-1 then are: SMC, UMi, Leo II, Ruprecht 106, Pal 12,
IC4499, NGC 6101, NGC 6934, Pal 4,
NGC 5053, NGC 5024 and NGC 5466, the only members farther than $2 Kpc$ from
the plane being Leo II and Pal 4.
\par
The radial velocities of the MP-1 candidate members, corrected for the motion
of the local standard of rest (LSR)
assuming a Galactic rotation of $220 Km s^{-1}$ and a solar motion
of $20 Kms^{-1}$ toward $l=57^0, b=22^0$ according to van den Bergh (1993),
are plotted in Figure 3 against the orbital longitudes calculated on the plane.
As can be seen, most of the candidate members are located on a {\it sinusoidal}
curve, so showing kinematical compatibility with a Keplerian motion.
The only evident exception is the highly retrograde cluster NGC 6934
which can thus be excluded from the group.
\par
Besides the above properties, the clue which mainly supports our
interest for this configuration is that {\it another of its members,
i.e. IC 4499, has been found by us to be a young globular cluster}
(see section 3.2).
\par
On the other hand, three of the quoted members, {\it i.e.} NGC 5053,
NGC 5024, and NGC 5466 are known to be very metal poor and old. Hence,
any claim of having detected a ``homogeneous group'' is not defendible,
or at least the degree of contamination by interlopers is quite large.
In this respect it is however also important to note that,
considering the available sample of about 20 clusters with reliable (relative)
TO-ages (Buonanno \etal 1995), a $\chi^2$-test shows  that the probability
that three ``young clusters'' appear to belong just by chance to the same
Great Circle in the sky is $\sim 11\%$, while such a probability for 3 ``old''
clusters is greater than $\sim70\%$.
\medskip
The Great Circle marked with a {\it dashed line} in Fig. 2 is the projection
on the Galactic sky of another ``curious'' plane (hereafter MP-2), which
is very similar to the Magellanic Plane presented in Fig. 1, but
accounts nicely also for the positions of Sgr dSph and of the two other
young globulars detected so far, {\it i.e.} Terzan 7 and Arp 2.
According to the criterium adopted above, the MP-2 candidate members
are: SMC, LMC, Arp 2, Terzan 7, NGC 2298, NGC 2808, NGC 6229, NGC 6715, NGC
2419, UMi, Dra, Car, and Sgr dSph. None of them is farther than $7 Kpc$
from MP-2.
Moreover, three clusters (NGC 1466, NGC 1841 and
ESO-121) whose membership to the Galaxy or LMC is uncertain lie very
close to the MP-2.
Admittedly, the statistical significance of this configuration
is lower than in MP-1 because {\it a)} only two young clusters are
presently members and {\it b)} Sgr dSph, Arp 2, Terzan 7, and NGC 6715 could
be  considered as a ``single object'' since (as noted by the Referee) ``they
fall on top of one another in partial phase space''.
\par
In Figure 4 we present the same plot as in Fig. 3 but for the MP-2 candidate
members. The points form again a clear sinusoidal pattern,
strongly suggestive of motion on Keplerian orbits. Note that
this configuration is very similar to that shown by
Kunkel (1979, Fig. 4) but it provides a remarkably better fit to a {\it sine}
curve and, in addition, it contains both Sgr dSph and the young
clusters Arp 2 and Terzan 7.
\par
As for MP-1, also in MP-2 there are objects which can be
excluded from a search for young cluster candidates. They are
NGC 2808, NGC 2298, and NGC 2419 which are known to be old from their Turn-Off
(TO) luminosities (Buonanno \etal 1989, VandenBerg \etal 1990,
Richer 1993). NGC 2419 is moreover the only point significantly deviating
from the overall pattern.
\par
The subgroup of MP-2 members which lie close to the Sgr dSph deserves
some more comments. Ibata \etal (1994) found that NGC 6715, Terzan 7 and
Arp 2 are physically neighbours of Sgr dSph and have rather similar
radial velocities, and on this basis suggested that they
could belong (or have originally belonged) to the Sgr dwarf galaxy.
The three clusters and the Sgr dSph are very nicely aligned
with respect to the centre of the Galaxy both in space and on the MP-2 plane.
Furthermore, following the method introduced by Kinman (1959),
we found that they all share clearly plunging orbits.
If the deep CCD photometric study
of NGC 6715 we are currently carrying on will show that also this cluster
is young (like Terzan 7 and Arp 2), the hypothesis of a common origin
for the whole sub-group would receive strong support given also the
preliminary age estimates obtained for the Sgr dSph, $\simeq 10 Gyr$
Mateo \etal (1994) (but see Sect. 3.3).
\par
Finally, for completeness, we mention that two more clusters could be
somehow associated to this group, {\it i.e.} Terzan 8 and Pal 8.
The available data on them are however still too uncertain to
get any reliable indication on their age and kinematics.
\medskip

\medskip\noindent
{\bf 3.2. Hunting for ``young'' globulars: IC 4499, a fortuitous alignment
or a confirmation?}

\medskip\noindent
IC 4499 is a low density southern  GGC frequently studied
in the past, but only our new photometric study (Ferraro \etal 1994) has
extended the photometry well below the TO-region, allowing for the first
time an estimate of the cluster age.
The mean cluster metallicity derived via different ``photometric'' indices
calibrated in terms of \fe ({\it i.e.} \bv0g, $\Delta V$, $S$,
$\delta(U-B)_{0.6}$, see Ferraro \etal 1994)
turns out to be \fe$=-1.75\pm0.20$ (the error is a conservative estimate),
which is lower by at least 0.2 $dex$ than the metallicity measured via other
{\it spectroscopic} techniques (see Tab. 1).
As stressed below, such a discrepancy has been noted
in {\it all} the young clusters detected so far.
\pn
In Figure 5a we present the CMD of IC 4499 containing 7217 stars
(variables are not plotted here) down to V$\sim 23$.
The measured magnitude difference between the turn-off and the horizontal
branch,
\dv$=3.25\pm0.15$, is significantly smaller than the mean value
of 3.55$\pm$0.09 found by Buonanno \etal (1989) for a sample of 18
well-studied GGCs.  This is a clear indication of younger age, and as
an example we
show in Figure 5b how the CMD ridge line of Arp 2 fits the
IC 4499 data. Since Arp 2 is one of the four ``young'' Galactic globulars
detected so far and has a ``photometric'' metallicity almost identical
to that measured for IC 4499, the excellent matching over the entire
CMD implies a very similar age, to within 1 Gyr or so.
In conclusion, it seems quite proven that  IC 4499 is another
``young'' globular cluster, displaying an age lower by 2-4 Gyr than ``normal''
clusters having similar metallicities.
\par
Therefore our search based on possible
spatial connections between previously known
young globulars has apparently been fruitful, since we have found another
young globular cluster at less than 2 Kpc
from the MP-1 and near the node between MP-1 and MP-2.

\medskip\noindent
{\bf 3.3. Chemical compositions and origin of the young globulars}

\medskip\noindent
There are essentially two items worth of consideration
in the chemical composition of the young clusters with respect to the
other normal clusters: a) the overall
metallicities, and b) the possible existence of common peculiarities.
\par
We already discussed the first item in a previous paper (Buonanno
\etal 1994a) noting that the young clusters display a very wide
metallicity range (from the metal-poor Ruprecht 106 and Arp 2 with
\fe $\sim -1.8$ up to the metal-rich Terzan 7 with \fe $=-0.49$,
Suntzeff \etal 1994).
This wide spread is undoubtedly a problem, and in particular it seems
difficult to explain how a metal-rich object like Terzan 7 could form
in a galactic system (Sgr dSph) having (presently) a quite low
mass (similar to the Fornax dSph) and with a mean metallicity
less than $-1$ (Ibata \etal 1994, Mateo \etal 1994).
\par
On the other hand, the characteristics of Terzan 7, which is
metal-rich, at $\sim15$ Kpc from the Galactic Centre and $\sim8$ Kpc from
the Galactic Plane, and with a plunging orbit (adopting the framework of
van den Bergh 1993), are hardly compatible also with a formation model of
a uniformly collapsing proto-Galaxy.
One could perhaps think of Terzan 7 as the ``nuclear'' remnant
of a self-enriched larger body now almost totally evaporated or disrupted.
However, the fact that its giant branch is so narrow (see Buonanno \etal
1994a,c) makes any self-enrichment unlikely.
Alternatively, it could have been
originated by a  High Velocity Cloud ejected from the disk of the Galaxy
by a supernova-driven ``Galactic fountain'' (Bregman 1980).
This hypothesis could be supported by the fact that old open clusters
located at similar galactocentric distances in the disk of the Galaxy
display metallicities similar to that of Terzan 7 (Friel 1993). Therefore
the cluster would now be infalling on a ballistic orbit.
A quite similar scenario could be envisaged also for Pal 8 if
the distance and height on the Galactic plane listed by Djorgovski
(1993) will be confirmed.
\medskip
The problem of having galaxies forming globulars more metal-rich
than the parent system could be reduced if the cluster metallicity
determinations were to be considered differently. In fact, as
noted in Sect. 3.2 for IC 4499, a problem apparently always present
in analysing the characteristics of the young clusters is that
the metallicity determinations obtained from ``photometric'' methods are
systematically lower (by about 0.2-0.3 dex) than the
spectroscopic determinations based on the CaII triplet
(Da Costa \etal 1992, Suntzeff \etal 1994) (see data in Table 1).
The problem is particularly evident for Terzan 7 where Buonanno
\etal (1994c) have determined a ``photometric'' metallicity
close to \fe$=-1.0$, much lower than the value obtained via ``spectroscopic''
means. If such a value were confirmed by further observations (or anyway,
if \fe were significantly smaller than the -0.49 obtained by Suntzeff \etal
1994), since the latest ``photometric'' estimate of the metallicity of
the Sgr dSph population is \fe$\sim -1.1\pm0.3$ (Mateo \etal 1994),
the hypothesis that Terzan 7 and the Sgr dSph could have a physical original
connection would get a significant support.
\par
The discrepancy between photometric and spectroscopic metallicity
determinations, apart from the noticeable case
of Terzan 7, is in general small enough to be accountable for by different
reasons in different clusters (e.g. photometric errors, reddening
uncertainties, calibration errors of the adopted photometric indices in terms
of \fe, etc.). However, we have investigated if there are intrinsic
physical reasons that might cause this discrepancy. One possibility is
the effect that a younger
age would have on the calibrations, but using the theoretical
models by the VandenBerg and the Frascati groups we find that this effect
is too small to explain the observed discrepancy (Buonanno
\etal 1993, 1994a,b,c, Ferraro \etal 1994).
Another possibility is a
different abundance ratio in the original material between the
typical Milky Way normal clusters and the young ones,
which would be in turn a signature of their origin from a different
environment. Observations of element ratios are not available yet for these
clusters, and are strongly urged for the potentially basic information they
may bring on this subject.
\medskip
As far as the structural and dynamical characteristics are concerned,
it is important to recall that: a)
all the young clusters detected so far are intrinsically smaller
and fainter than the average globular clusters in the Galaxy and, as remarked
by van den Bergh (1994b), they show structural characteristics similar to
those displayed by the halo clusters of the Magellanic Clouds and of the
Fornax dSph; and b)
all the five known young globulars (with perhaps
the exception of Pal 12) are on {\it plunging} orbits, and two
of them ({\it i.e.} Ruprecht 106 and IC4499) have prograde rotation,
according to the method adopted by Kinman (1959) and van den Bergh (1993).
\par
In conclusion, the origin of the young GGCs (the metal-rich ones,
in particular) is not yet understood, but there are increasing hints
on a possible connection to episodes of galaxy interactions.

\medskip\noindent
{\bf 3.4 The Andromeda Plane and nodes: just a curiosity?}

\medskip\noindent
Before closing this section, it may be interesting to point out another
plane whose existence has emerged quite clearly
from our tests. This new plane, which we call hereafter {\it the
Andromeda Plane}, roughly accounts for the Galactic centre, M31 and its
satellites and the
Magellanic Clouds. As can be seen from the plot in Figure 6, it yields also
a fairly good alignment with M33, IC 10 and IC 1613, and, obviously
Draco and Ursa Minor.
Note that $50 \%$ of the certain Local Group members lie within
$18 Kpc$ from this plane. Furthermore, a Kolmogorov-Smirnov test rejects
the hypothesis that the considered sample is drawn from a parent population
randomly distributed around the plane with a more than $99 \%$ confidence
level even considering the zone of galactic obscuration, according to
Kunkel (1979).
\par
Though it is surely too simplistic to imagine a description of the
Local Group galaxies as populating just two or three fundamental planes,
it is on the other hand at least puzzling that
most (if not {\it all}) the galaxies known so far to be members of the
Local Group actually lie in or very close to these planes and their
respective nodes. This consideration could acquire special relevance
if one recalls that several authors (e.g. Valtonen \etal 1993,
Byrd \etal 1994 and Lynden-Bell 1993) have already speculated
about the occurrence of major interactions between the main members
of the Local
Group also involving physical complanarities between the Galaxy, M31 and the
neighbours non-LG-members IC 342 and Maffei 1,
and the possible capture of the MCs
from M31.

\bigskip\noindent
{\bf 4. ABOUT THE DEFINITION OF ``YOUNG'' AND ITS IMPLICATIONS}

\medskip\noindent
In recent years, there has been a growing consensus on age being
the {\it ``second parameter''} in the outer regions
of the Galaxy (Zinn 1993). Though there is, in our view, strong evidence
that age cannot be {\it the only} second parameter at work
(see for instance Fusi Pecci \etal 1993, Rood \etal 1993, Catelan and de
Freitas Pacheco 1994,  Buonanno 1994 and references therein),
the conclusion that
the {\it global} formation phase of the Galactic halo lasted longer than
originally proposed by Eggen, Lynden-Bell and Sandage (ELS, 1962)
or by Yoshii and Saio (1984) seems consistent
with the latest observational data and hardly avoidable (Majewski 1993a).
Therefore we may assume that in the early stages of the Galaxy
formation there has been a rapid ($\simeq 1 Gyr$)
collapse with accompanying chemical evolution of a large protogalactic cloud
producing the bulk of the old halo and disk clusters, followed by a (chaotic)
merging
over several Gyrs of multiple fragments that experienced independent chemical
evolution. The attention is now mostly focused on
identifying which clusters (or group of clusters) belong to a given
population or to a given original ``fragment'' (Searle and Zinn 1978).
\par
Following the very valuable approaches of Zinn and collaborators and
van den Bergh, the basic tools used so far for this purpose are essentially the
HB morphology as age indicator, and the present Galactocentric location and
radial velocity as indicators of the kinematical properties. Several studies
(see in particular Rodgers and Paltoglou 1984, Zinn 1993,
van den Bergh 1993, 1994b, Majewski 1994) have also led to identify
specific sub-groups of clusters which could share common original
properties. However, it is important to note that the definition of ``young''
based on the HB morphology may be ambiguous if age is not {\it the only}
second parameter driving the star distribution along the HB. Also
the true kinematical properties are actually unknown until
complete space velocities will be precisely measured.
\par
In this sense, our definition of ``young'' globular cluster is different
from that used for instance in the Zinn's terminology, as we rest on
{\it the actual measure of the TO luminosity}. The reason why we think
this note is important can be seen, for instance, with respect to the very
interesting results presented by Majewski (1994).\pn
Very schematically, Majewski (1994) reports that the group of the 10-13
reddest HB, young halo clusters in the Zinn list appear to populate
the FLS plane suggesting a scenario where the FLS stream and the red-HB
young halo globular clusters share a common origin from
a large fragment {\it \`a la} Searle and Zinn or from a former
parent satellite galaxy.
Based on the list presented in Table 1 of Majewski (1994) one may notice
however, that the 10 clusters with the HB parameter
taken from Lee \etal (1993) \lz$<-0.5$ ({\it i.e.} with red HBs)
display a wide range of {\it relative} ages when the MS and TO
regions are considered. In particular, Pal 12 and Ruprecht 106
are surely younger than normal clusters by 3-4 Gyr based
on their TO properties, whereas NGC 1261 and 2808 have
normal ages (at least as far as the available TO data indicate,
see Bolte and Marleau 1989, Buonanno \etal 1983, respectively).
Moreover, they span a very wide range in
Galactocentric distances ranging from $\sim 8$ to $\sim 110$ Kpc,
and quite different types of orbits (at least in the framework of
van den Bergh 1993). If one extends the sample to the 13 clusters
with \lz$<-0.3$, the situation is even worse as one would mix
clusters with known prograde and retrograde orbits.
Therefore, given the spread of observed properties,
a ``single-event'' common origin for this  group
seems quite unlikely, as it would require a fairly large fragment with
very special conditions of orbit decaying.
The internal inconsistencies of the scenario which makes use
of any HB parameter as age indicator was pointed out also in the
Majewski's paper (Fig. 4 and note 5), where
he finds that the clusters displaying the largest
$\Delta$\lz ({\it i.e.} the parameter more directly correlated to
age differences, Zinn 1993) do not correlate with the FLS plane
as significantly as the red-HB clusters do.
\par
On the other hand, though necessarily not very significant on statistical
grounds given the very small number of involved objects, interesting hints
can be found considering subsets of the clusters singled out by Majewski (1994)
or others. For instance, Pal 4, Pal 14, Eridanus, and (slightly
worse) Pal 3 lie in a plane including the Galactic Centre, are very distant
(far beyond the $33 - 60$ Kpc radial gap noted by Zinn (1985) in the
GGC distribution), are faint and morphologically similar,
span a small metallicity range, lie all on plunging orbits.
So, for what our analysis is concerned, the major hints about
``common-origin groups'' favours small scale phenomena rather than the
single-event accretion of large fragments, comprehensive of $8-10$ globulars,
as envisaged by Majewski (1994) or van den Bergh (1993).
\medskip
In conclusion, though the various claims (including our own) of possible
families of globular clusters  born from a common proto-fragment are
attractive and worthy of further detailed studies, it may be wise to wait
for an extended survey of the TO-properties of the considered globulars,
at least until the second parameter problem is definitely settled,
before accepting oversimplified models of Galaxy formation.
Furthermore, it is worth noting that different proposed scenarios
(for example recent capture from still existing systems, primordial
merging of self-enriched fragments, or formation induced by close
dynamical interactions between galaxies) are not equivalent and none
of them can be presently ruled out on observational basis.
Crucial in this respect is the determination of
mass of the fragment(s)/galaxy which eventually merged into the Milky Way,
as this will pose very strong constraints on the formation/destruction
of the thin and thick disks of our own Galaxy (Quinn \etal 1993).
In our view, there is presently no clear-cut observational evidence
against a model of Galaxy formation of the type ``main ~collapse''
{\it \`a la} ELS $+$ ``noise'' {\it \`a la} SZ, provided that
a slightly longer timescale for the main collapse event
($\sim1-2$ Gyr, Sandage 1993) is adopted.

\bigskip\noindent
{\bf 5. CONCLUDING REMARKS}

\medskip\noindent
The basic aim underlying the present study is the definition of a
procedure able to yield, first, possible candidate
{\it young} globular clusters and, second, a global scenario
which could somehow link the origin of these young clusters to
past interactions between the Milky Way and its satellites.

This task has been carried out by looking for spatial configurations
involving both clusters and galaxies and following the
approach used in the early '70s for instance by Kunkel, Demers and
Lynden-Bell to suggest the existence of a few Great Circles (the
Magellanic Plane and the Fornax-Leo-Sculptor Plane) which ideally
contain various (GGCs and galaxies) satellites of the Milky Way.

The main conclusions of the present report are:

\pn
\item{1.} We present two planar configurations which link the {\it
young} globular clusters detected so far in the Milky Way to its
satellites. In particular,  one of them --MP-1--  connects Pal 12
and Ruprecht 106 to SMC and The Magellanic Stream, supporting
the hypothesis of a "magellanic origin" for these two cluster
(Lin and Richer 1992).

\pn
\item{2.} Using the present approach, we found another candidate
{\it young} globular cluster, IC 4499, lying on MP-1: so three over
five of the presently known young Galactic globulars  are apparently
aligned on MP-1 and display kinematical properties compatible with
that of other MP-1 members. This cannot be taken yet as a proof of
(hystorical) physical connections, but may be worth of further
study.

\pn
\item{3.} While discussing about the detection and origin of the
{\it young} globular clusters, we present some warnings and comments
about the need for a clear-cut definition of the term ``young''
here based on the Turnoff properties whilst it is generally used also
adopting the Horizontal Branch morphology as a safe age-indicator (Lee
1993, Zinn 1993). In particular,
we show that this may lead to rather unplausible identifications of
subgroups of clusters whose common origin is envisaged.

\bigskip\noindent
{\bf Acknowledgements:}
We warmly thank Roberto Buonanno, Carlo Corsi, Ivan Ferraro, Harvey
Richer, and Greg Fahlman, members of the group involved in the whole
project, for the lively collaboration and discussions. We also thank
Paolo Montegriffo for the precious help in developing the codes used for the
analysis. We are indebted to the Referee for helpful comments
and suggestions.
This work was supported by the {\it Consiglio per le
Ricerche Astronomiche} of the {\it Ministero per l' Universit\`a
e la Ricerca Scientifica e Tecnologica}.

\bigskip\bigskip\bigskip
\parskip=0pt
\centerline{\bigbf References}
\bigskip
\pp Armandroff, T. E., 1989, AJ, 97, 371

\pp Armandroff, T. E. \& Da Costa, G. S., 1991, AJ, 101, 1329 (AD91)

\pp Bolte, M. \& Marleau, F. 1989, PASP, 101, 1088

\pp Bregman, J. N. 1980, ApJ, 236, 571

\pp Buonanno, R. 1994, in The Formation of The Milky Way, eds. E. Alfaro
and G. Tenorio Tagle (Cambridge University Press), (in press).

\pp Buonanno, R., Corsi, C.E. \& Fusi Pecci, F. 1989, A\&A, 216, 80

\pp Buonanno, R., Corsi, C.E., Fusi Pecci, F. \& Harris, W.E. 1983,
AJ, 89, 365

\pp Buonanno, R., Buscema, G., Fusi Pecci, F., Richer, H. B. \& Fahlman, G. G.
1990, AJ, 100, 1811 (B90)

\pp Buonanno, R., Corsi, C. E., Fusi Pecci, F., Richer, H. B. \& Fahlman, G. G.
1993, AJ, 105, 181

\pp Buonanno, R., Corsi, C. E., Fusi Pecci, F., Richer, H. B. \& Fahlman, G. G.
1994a, ApJ, 430, L121

\pp Buonanno, R., Corsi, C. E., Fusi Pecci, F., Richer, H. B. \& Fahlman, G. G.
1994b, AJ, (in press) (B94b)

\pp Buonanno, R., Corsi, C. E., Fusi Pecci, F., Richer, H. B. \& Fahlman, G. G.
1994c, AJ, (in press) (B94c)

\pp Buonanno, R., Corsi, C. E., Pulone, L., Bellazzini, M.,
Ferraro, F.R., \& Fusi Pecci, F., 1995, (in preparation).

\pp Byrd, G., Valtonen, M., McCall, M. \& Innanen, K. 1994, AJ, 107, 2051

\pp Catelan, M. \& de Freitas Pacheco, J. A. 1994, A\&A, 289, 394

\pp Da Costa, G. S. \& Armandroff, T. E. 1990, AJ, 100, 162 (DA90)

\pp Da Costa, G. S., Armandroff \& T. E., Norris, J. E. 1992, AJ, 104, 154
(DAN)

\pp Djorgovski, S. G., 1993, in Structure and Dynamics of Globular Clusters,
eds. S. G. Djorgovski and G. Meylan (ASP S.Francisco),
ASP Conf. Ser., No. 50, p.371

\pp Eggen, O. J., Lynden-Bell, D. \& Sandage, A. 1962, ApJ, 136, 738 (ELS)

\pp Ferraro, I., Ferraro, F.R., Fusi Pecci, F., Corsi, C. E. \& Buonanno, R.
1994, MNRAS, (submitted) (F94)

\pp Freeman, K. C., Illingworth, G. \& Oemler A. Jr. 1983, ApJ, 272, 488

\pp Friel, E. D., 1993, in The Globular Cluster-Galaxy Connection,
eds. G. H. Smith and J. P. Brodie (ASP S.Francisco),
ASP Conf. Ser., No. 48, p.271

\pp Fusi Pecci, F., Ferraro, F. R., Bellazzini, M., Djorgovski, G. S., Piotto,
G. \& Buonanno, R. 1993, AJ, 105, 1145

\pp Gratton, R. \& Ortolani, S., 1988, A\&AS, 73, 131

\pp Hernquist, L. \& Bolte, M. 1993, in The Globular Cluster-Galaxy Connection,
eds. G. H. Smith and J. P. Brodie (ASP S.Francisco), ASP Conf. Ser.,
No. 48, p.781

\pp Hodge, P. W. \& Michie R. W. 1969, AJ, 74, 581

\pp Ibata, R. A., Irwin, M. J. \& Gilmore, G. 1994, Nature, 370, 194

\pp Johnston, K. V. \& Hernquist, L., 1993, in Proceedings of the ESO/OHP
Workshop on Dwarf Galaxies, G. Meylan and P. Prugniel eds., p. 389

\pp Johnston, K. V., Spergel, D. N. \& Hernquist, L., 1995, Preprint

\pp Kinman, T. D. 1959, MNRAS, 119, 551

\pp Kunkel, W. E. 1979, ApJ, 228, 711

\pp Kunkel, W. E. \& Demers, S. 1977, ApJ, 214, 21

\pp Lee, Y. W., Demarque, P. \& Zinn, R., 1993, ApJ, 423, 241

\pp Lin, D. N. C. \& Richer, H. B. 1992, ApJ, 388, L51

\pp Lynden-Bell, D. 1976, MNRAS, 174, 691

\pp Lynden-Bell, D. 1982, The Observatory, 102, 201

\pp Lynden-Bell, D. 1993, in Proceedings of the ESO/OHP Workshop on
Dwarf Galaxies, G. Meylan and P. Prugniel eds., p. 589

\pp Majewski, S. R. 1993a, ARA\&A, 31, 571

\pp Majewski, S. R., 1993b, in Galaxy Evolution: The Milky Way Perspective,
ed. S. R. Majewski (ASP, S. Francisco)
ASP Conf. Ser., No. 49, 1

\pp Majewski, S.R. 1994, ApJ, 431, L17

\pp Mateo, M., Udalski, A., Szymanski, M., Kaluzny, J., Kubiak, M. \&
Krzeminski, W. 1994, (preprint)

\pp Peterson, C. J., 1993, in Structure and Dynamics of Globular Clusters,
eds. S. G. Djorgovski and G. Meylan (ASP S.Francisco),
ASP Conf. Ser., No. 50, p.331

\pp Piatek, S. \& Pryor, C., 1995, AJ, 109, 1071

\pp Quinn, P.J., Hernquist, L. \& Fullagar, D.P. 1993, ApJ, 403, 71

\pp Richer, H.B. 1993, in The Globular Cluster-Galaxy Connection,
eds. G. H. Smith \& J. P. Brodie (ASP, S.Francisco),
ASP Conf. Ser., 48 (presented in the panel discussion).

\pp Rodgers, A. W. \& Paltoglou, G., 1984, ApJ, 283, L1

\pp Rood, R. T., Crocker, D. A., Fusi Pecci, F., Ferraro, F. R., Clementini,
G. \& Buonanno, R. 1993, in The Globular Cluster-Galaxy Connection,
eds. G. H. Smith \& J. P. Brodie (ASP, S.Francisco),
ASP Conf. Ser., 48, p.218

\pp Sandage, A.R. 1993, AJ, 106, 719

\pp Sarajedini, A. 1994, AJ, 107, 618 (Sa94)

\pp Searle, L. \& Zinn, R. 1978, Ap.J, 225, 351

\pp Smith, H. A. 1984, ApJ, 281, 148 (Sm84)

\pp Stetson, P. B., Vandenbergh, D. A., Bolte, M. A., Hesser, J. E. \& Smith,
G. H. 1989, AJ, 97, 1360

\pp Suntzeff, N., Richer, H. B. \& Lin, D. N. C., 1994, (preprint) (S94)

\pp Thomas, P. 1989, MNRAS, 238, 1311

\pp Valtonen, M. J., Byrd, G. G., McCall, M. L. \& Innanen, K. A. 1993, AJ,
 105, 881

\pp van den Bergh, S. 1993, AJ, 105, 971

\pp van den Bergh, S. 1994a, AJ, 107,1328

\pp van den Bergh, S. 1994b, (preprint)

\pp VandenBerg, D.A., Bolte, M., \& Stetson, P.B. 1990, AJ, 100, 445

\pp Velazquez, H. \& White, D. M., 1995, {\it to appear in} MNRAS
({\it Letters})

\pp Webbink, R. 1985, in Dynamics of Star Clusters, IAU Symp. n.113,
eds. J. Goodman and P. Hut (Reidel, Dordrecht), p.541 (W85)

\pp Yoshii, Y. \& Saio, H. 1979, PASJ, 31, 331

\pp Zaritsky, D., 1994, (preprint)

\pp Zinn, R. 1985, ApJ, 293, 421

\pp Zinn, R., 1993 in The Globular Cluster-Galaxy Connection,
eds. G. H. Smith \& J. P. Brodie (ASP, S.Francisco),
ASP Conf. Ser., No. 48, 31

\pp Zinn, R. \& West, M. J. 1984, ApJ, 55, 41 (ZW84)

\vfill\eject

\parskip=5pt
\centerline{\bigbf Figure Captions:}
\bigskip\medskip

\noindent{\bf Figure 1:}{ The Galactocentric Great Circles representing
the Magellanic Plane ({\it MP, full line}) and the
Fo\-r\-n\-ax-Leo-Scul\-p\-t\-or Pla\-ne ({\it FLS, dot\-ted li\-ne}),
re\-s\-pec\-ti\-ve\-ly.
The coordinates are Galactocentric Galactic longitude
and latitude, in degrees. Labels are abbreviated names or NGC numbers.
The patterns at the low corners of
the plot are schematic representations of the Magellanic Stream according to
Kunkel (1979).}

\bigskip
\noindent{\bf Figure 2:}{ The MP-1 ({\it dashed line}) and
MP-2 ({\it full line}) Galactocentric Great Circles (see Sect. 3.1)
in the same coordinate system as in Fig. 1. The planes are
constrained to pass through the Galactic Centre. Note that the MP-2
nearly fits the extension of the Magellanic Stream.}

\bigskip
\noindent {\bf Figure 3:}{ Radial velocities of MP-1 members, corrected
for both LSR and solar motions, as a function of the orbital longitude. The
{\it full line} represents a first order fit to the data of the Equation 1
of Kunkel (1979). }

\bigskip
\noindent {\bf Figure 4:}{ Radial velocities of MP-2 members, corrected
for both LSR and solar motions, as a function of the orbital longitude. The
{\it full line} represents a first order fit to the data of the Equation 1
of Kunkel (1979). }

\bigskip
\noindent{\bf Figure 5:}{ {\it a)} The Color Magnitude Diagram of
the Galactic globular cluster IC 4499, lying on the MP-1;
{\it b)} the mean ridge line of Arp 2 superposed on the IC 4499
data. As can be deduced from the nice fit in the Turnoff region,
the age of the two clusters is the same to within 1 Gyr (see Sect. 3.2).
}

\bigskip
\noindent{\bf Figure 6:}{ The {\it Andromeda Plane} (see Sect. 3.4)
plotted in the same coordinate system as in Fig. 1 and 2.
Note that $\sim50\%$ of the galaxies which are certain members of
the Local Group are located within $18 Kpc$ from this plane.
A zoomed plot of the region surrounding M31 is also presented: the alignement
of galaxies in this region is rather striking.}

\vfill\eject

\def\page{\footline={\ifnum\pageno=1 \hfil
          \else\hss\tenrm\folio\hss\fi}}

\def\ahead#1\smallskip{{\vfil\eject{\centerline{\bf#1}}
\smallskip}{\message#1}}
\def\bhead#1\smallskip{{\bigbreak{\centerline{\it#1}}\smallskip}
{\message#1}}
\def\chead#1\par{{\bigbreak\noindent{\it#1}\par} {\message#1}}

\def\head#1. #2\par{\medbreak\centerline{{\bf#1.\enspace}{\it#2}}
\par\medbreak}

\def\levelone#1~ ~ #2\smallskip{\noindent#1~ ~ {\bf#2}\smallskip}
\def\leveltwo#1~ ~ #2\smallskip{\noindent#1~ ~ {\it#2}\smallskip}
\def\levelthree#1~ ~ #2\smallskip{\noindent#1~ ~ {#2}\smallskip}

\def\m{^m\kern-7pt .\kern+3.5pt}
\def\p{^{\prime\prime}\kern-2.1mm .\kern+.6mm}
\def\pone{^{\prime}\kern-1.05mm .\kern+.3mm}
\def\dpoint{^d\kern-1.05mm .\kern+.3mm}
\def\hpoint{^h\kern-2.1mm .\kern+.6mm}
\def\y{^y\kern-1.05mm .\kern+.3mm}
\def\s{^s\kern-1.2mm .\kern+.3mm}

\def\apgt{\ {\raise-.5ex\hbox{$\buildrel>\over\sim$}}\ }
\def\aplt{\ {\raise-.5ex\hbox{$\buildrel<\over\sim$}}\ }
\def\deg{^{\circ}}

\def\hup{^{h}\kern-2.1mm .\kern+.6mm}

%
%

%
%

%
%
\def\today{\number\day\space\ifcase\month\or
January\or February\or March\or April\or May\or June\or July\or
August\or September\or October\or November\or December\fi
\space\number\year}

\def\sqr#1#2{{\vcenter{\hrule height.#2pt
\hbox{\vrule width.#2pt height#1pt \kern#1pt
\vrule width.#2pt}
\hrule height.#2pt}}}

\newcount\equationnumber
\newbox\eqprefix
\def\neweqprefix#1{\global\equationnumber=1
\global\setbox\eqprefix=\hbox{#1}}

\def\autono{(\copy\eqprefix\number\equationnumber)
\global\advance\equationnumber by 1}

\def\trule{\vskip6pt\hrule\vskip2pt\hrule\vskip6pt}
\def\mrule{\noalign{\vskip6pt\hrule\vskip6pt}}

%
%
%
%

\def\etal{{\it et~al.\ }}

\newcount\refno
\refno=0

\def\beginorefs\par{\begingroup\parindent=12pt
\frenchspacing \parskip=1pt plus 1pt minus 1pt
\interlinepenalty=1000 \tolerance=400 \hyphenpenalty=10000
\everypar={\item{\the\refno.}\hangindent=2.6pc}

\def\nature##1,{{\it Nature}, {\bf##1},}

\def\aa##1,{{\it Astr.\ Ap.,\ }{\bf##1},}
\def\aapr{{\it Astr.\ Ap.,\ }in press.}
\def\ajaa##1,{{\it Astron.\ Astrophys.,\ }{\bf##1},}
\def\ajaapr{{\it Astron.\ Astrophys.,\ }in press.}

\def\aalet##1,{{\it Astr.\ Ap.\ (Letters),\ }{\bf##1},}
\def\aaletpr{{\it Astr.\ Ap.\ (Letters),\ }in press.}
\def\ajaalet##1,{{\it Astron. Astrophys. (Letters),\ }{\bf##1},}
\def\ajaaletpr{{\it Astron. Astrophys. (Letters),} in press.}

\def\aasup##1,{{\it Astr. Ap. Suppl.,\ }{\bf##1},}
\def\aasuppr{{\it Astr.\ Ap.\ Suppl.,\ }in press.}
\def\ajaasup##1,{{\it Astron. Astrophys. Suppl.,\ }{\bf##1},}
\def\ajaasuppr{{\it Astron.\ Astrophys.\ Suppl.,\ }in press.}

\def\aass##1,{{\it Astr. Ap. Suppl. Ser.,\ }{\bf##1},}
\def\aasspr{{\it Astr. Ap. Suppl. Ser.,} in press.}

\def\aj##1,{{\it A.~J.,\ }{\bf##1},}
\def\ajpr{{\it A.~J.,\ }in press.}
\def\ajaj##1,{{\it Astron.~J.,\ }{\bf##1},}
\def\ajajpr{{\it Astron.~J.,} in press.}

\def\apj##1,{{\it Ap.~J.,\  }{\bf##1},}
\def\apjpr{{\it Ap.~J.,} in press.}
\def\ajapj##1,{{\it Astrophys.~J.,\ }{\bf##1},}
\def\ajapjpr{{\it Astrophys.~J.,} in press.}

\def\apjlet##1,{{\it Ap.~J. (Letters),\ }{\bf##1},}
\def\apjletpr{{\it Ap.~J. (Letters),} in press.}
\def\ajapjlet##1,{{\it Astrophys. J. Lett.,\ }{\bf##1},}
\def\ajapjletpr{{\it Astrophys. J. Lett.,} in press.}

\def\apjsup##1,{{\it Ap.~J.~Suppl.,\ }{\bf##1},}
\def\apjsuppr{{\it Ap.~J.\ Suppl.,} in press.}
\def\ajapjsup##1,{{\it Astrophys. J. Suppl.,\ }{\bf##1},}
\def\ajapjsuppr{{\it Astrophys. J.\ Suppl.,} in press.}

\def\araa##1,{{\it Ann. Rev. A.~A.,\ }{\bf##1},}
\def\araapr {{\it Ann. Rev. A.~A.,} in press.}

\def\baas##1,{{\it B.A.A.S.,\ }{\bf##1},}
\def\baaspr{{\it B.A.A.S.,} in press.}

\def\mnras##1,{{\it M.N.R.A.S.,\ }{\bf##1},}
\def\mnraspr{{\it M.N.R.A.S.,} in press.}
\def\ajmnras##1,{{\it Mon. Not. R. Astron. Soc.,\ }{\bf##1},}
\def\ajmnraspr{{\it Mon. Not. R. Astron. Soc.,} in press.}

\def\pasp##1,{{\it Pub.~A.S.P.,\ }{\bf##1},}
\def\pasppr{{\it Pub.~A.S.P.,} in press.}
\def\ajpasp##1,{{\it Publ. Astron. Soc. Pac.,\ }{\bf##1},}
\def\ajpasppr{{\it Publ. Astron. Soc. Pac.,} in press.}
}

\def\eighteenpoint{
  \font\eighteeni=cmmi10 scaled\magstep3
  \font\eighteensy=cmsy10 scaled\magstep3
  \font\eighteenrm=cmr10 scaled\magstep3
  \font\twelvei=cmmi12
  \font\twelvesy=cmsy12
  \font\teni=cmmi10
  \font\tensy=cmsy10
  \font\seveni=cmmi7
  \font\sevensy=cmsy7
  \font\it=cmti10 scaled \magstep3
  \font\bf=cmb10 scaled \magstep3
  \font\sl=cmsl10 scaled \magstep3
  \textfont0= \eighteenrm \scriptfont0=\twelverm
\scriptscriptfont0=\tenrm
  \def\rm{\fam0 \eighteenrm}
  \textfont1=\eighteeni  \scriptfont1=\twelvei
\scriptscriptfont1=\teni
  \def\mit{\fam1 } \def\oldstyle{\fam1 \eighteeni}
  \textfont2=\eighteensy \scriptfont2=\tensy
\scriptscriptfont2=\sevensy
\def\doublespace{\baselineskip=30pt\lineskip=0pt
\lineskiplimit=-5pt}
\def\singlespace{\baselineskip=20pt\lineskip=0pt
\lineskiplimit=-5pt}
\def\oneandahalf{\baselineskip=25pt\lineskip=0pt
\lineskiplimit=-5pt}
\def\deg{^{\raise2pt\hbox{$\circ$}}}}


\newdimen\digitwidth
\setbox0=\hbox{\rm0}
\digitwidth=\wd0

\vsize=18truecm
\hsize=16truecm
\nopagenumbers
\voffset=-6truemm
\tabskip=2em plus1em minus1em
{\bf Table 1.} Metallicities of the Young G.C.: spectroscopic vs. photometric.
\trule
\halign to
\hsize{
\hfil#\hfil&
\hfil#\hfil&
\hfil#\hfil&
\hfil#\hfil&
\hfil#\hfil\cr
 $Name$ &
$ [Fe/H]_{phot}$ &
$ Ref $ &
$ [Fe/H]_{spec} (CaII)$ &
$ Ref $ \cr
\mrule
Terzan~7&$-1.00\pm 0.13$&B94c&$-0.49\pm 0.05$&S94~ \cr
{}~~~~~~~~&~~~~~~~~~~~~~~~&~~~~&~~~~~~~~~~~~~~~&~~~~ \cr
{}~~~~~~~~&~~~~~~~~~~~~~~~&~~~~&~~~~~~~~~~~~~~~&~~~~ \cr
IC~4499~&$-1.75\pm 0.20$&F94~&$-1.5~\pm 0.3$~&ZW84 \cr
{}~~~~~~~~&$-1.77\pm 0.20$&W85~&$-1.38\pm 0.2$~&Sm84 \cr
{}~~~~~~~~&~~~~~~~~~~~~~~~&~~~~&~~~~~~~~~~~~~~~&~~~~ \cr
Rup~106~&$-1.90\pm 0.20$&B90~&$-1.69\pm 0.05$&DAN~ \cr
{}~~~~~~~~&$-1.85\pm 0.20$&W85~&~~~~~~~~~~~~~&~~~~ \cr
{}~~~~~~~~&$-1.61\pm 0.20$&Sa94&~~~~~~~~~~~~~~~&~~~~ \cr
{}~~~~~~~~&~~~~~~~~~~~~~~~&~~~~&~~~~~~~~~~~~~~~&~~~~ \cr
Arp~2~~~&$-1.84\pm 0.25$&B94b&$-1.73\pm 0.05$&S94~ \cr
{}~~~~~~~~&$-1.85\pm 0.20$&W85~&~~~~~~~~~~~~~&~~~~ \cr
{}~~~~~~~~&~~~~~~~~~~~~~~~&~~~~&~~~~~~~~~~~~~~~&~~~~ \cr
Pal~12~~&$-1.13\pm 0.20$&W85~&$-0.60\pm 0.11$&AD91 \cr
{}~~~~~~~~&$-1.06\pm 0.12$&DA90&$-1.0\pm0.1$~&AD91 \cr
{}~~~~~~~~&$-0.98\pm 0.14$&Sa94&~~~~~~~~~~~~~~~&~~~~ \cr
}\trule
\noindent
AD91 = Armandroff \& Da Costa 1991;
B90 = Buonanno {\it et. al.} 1990;
B94b = Buonanno {\it et. al.} 1994b;
B94c = Buonanno {\it et. al.} 1994c;
DA90 = Da Costa \& Armandroff 1990;
DAN = Da Costa, Armandroff \& Norris 1992;
F94 = Ferraro {\it et. al.} 1994;
Sa94 = Sarajedini 1994;
S94 = Suntzeff {\it et. al.} 1994;
Sm84 = Smith 1984;
W85 = Webbink 1995;
ZW84 = Zinn \& West 1984.
\vfill\eject
\bye